\def\spose#1{\hbox to 0pt{#1\hss}}
\def\approxlt{\mathrel{\spose{\lower 3pt\hbox{$\sim$}}
        \raise 2.0pt\hbox{$<$}}}
\def\approxgt{\mathrel{\spose{\lower 3pt\hbox{$\sim$}}
        \raise 2.0pt\hbox{$>$}}}

\def\multleft#1{\hbox to size{\vbox {\halign {\lft{##}\cr #1}}\hfill}\par}
\def\multright#1{\hbox to size{\vbox {\halign {\rt{##}\cr #1}}\hfill}\par}

\def\degmark{^\circ}
\def\boxit#1{\vbox{\hrule\hbox{\vrule\kern3pt\vbox{\kern3pt
          #1 \kern3pt}\kern3pt\vrule}\hrule}}

\def\cm{{\rm\thinspace cm}}

\def\erg{{\rm\thinspace erg}}
\def\eV{{\rm\thinspace eV}}

\def\K{{\rm\thinspace K}}
\def\keV{{\rm\thinspace keV}}
\def\km{{\rm\thinspace km}}

\def\Mpc{{\rm\thinspace Mpc}}
\def\Msun{\hbox{$\rm\thinspace M_{\odot}$}}
\def\pc{{\rm\thinspace pc}}
\def\ph{{\rm\thinspace ph}}
\def\s{{\rm\thinspace s}}
\def\yr{{\rm\thinspace yr}}

\def\pcmcu{\hbox{$\cm^{-3}\,$}}

\def\ergps{\hbox{$\erg\s^{-1}\,$}}

\def\kmps{\hbox{$\km\s^{-1}\,$}}

\def\pcmsq{\hbox{$\cm^{-2}\,$}}

\def\phpcmsqpspkeV{\hbox{$\ph\cm^{-2}\s^{-1}\keV^{-1}\,$}}

\def\ps{\hbox{$\s^{-1}\,$}}

\def\pyr{\hbox{$\yr^{-1}\,$}}
\def\kmpspMpc{\hbox{$\kmps\Mpc^{-1}$}}

\def\pct{\%}
\def\mcg6{MCG--6-30-15}

\def\Msunpyr{\hbox{$\Msun\pyr$}}

\documentclass[]{emulateapj}
\usepackage{natbib}
\shorttitle{Chandra HETGS Spectral Study of the Iron K Bandpass in \mcg6}

\shortauthors{A. J. Young, J. C. Lee, A. C. Fabian, C. S. Reynolds, R. R. Gibson \& C. R. Canizares}

\begin{document}

\title{A Chandra HETGS Spectral Study of the Iron K Bandpass in \mcg6:\\  A Narrow View of the Broad Iron Line}

\author{A. J. Young\footnotemark[1], J. C. Lee\footnotemark[2,3], A. C. Fabian\footnotemark[4], C. S. Reynolds\footnotemark[5], R. R. Gibson\footnotemark[1], C. R. Canizares\footnotemark[1].}

\footnotetext[1]{MIT Center for Space Research, 77 Massachusetts Avenue, Cambridge, MA 02139}

\footnotetext[2]{Harvard-Smithsonian Center for Astrophysics, 60 Garden Street MS-4, Cambridge, MA 02138}

\footnotetext[3]{Chandra fellow}

\footnotetext[4]{Institute of Astronomy, University of Cambridge, Madingley Road, Cambridge, CB3 0HA, UK}

\footnotetext[5]{Department of Astronomy, University of Maryland, College Park, MD 20742}

% Notice that each of these authors has alternate affiliations, which
% are identified by the \altaffilmark after each name.  The actual alternate
% affiliation information is typeset in footnotes at the bottom of the
% first page, and the text itself is specified in \altaffiltext commands.
% There is a separate \altaffiltext for each alternate affiliation
% indicated above.

% The abstract environment prints out the receipt and acceptance dates
% if they are relevant for the journal style.  For the aasms style, they
% will print out as horizontal rules for the editorial staff to type
% on, so long as the author does not include \received and \accepted
% commands.  This should not be done, since \received and \accepted dates
% are not known to the author.

\begin{abstract}

We present a high resolution X-ray spectrum of the iron K bandpass in \mcg6 based on a 522~ksec observation with Chandra's High Energy Transmission Grating Spectrometer.  The Chandra spectrum is consistent with the presence of a relativistically broadened, highly redshifted iron K$\alpha$ emission line with a similar profile to previous observations.  A number of narrow features are detected above 2 keV, including a narrow Fe K$\alpha$ emission line and narrow absorption lines from H- and He-like Fe, H-like S and H-like Si.  This absorption is well described by a photoionized plasma with a column density $\log N_H = 23.2$ and an ionization parameter $\log \xi = 3.6$, assuming the iron abundance has the Solar value and a velocity dispersion parameter $b = 100 \kmps$.  Applying this absorption model to a high fidelity XMM-Newton EPIC-pn spectrum we find that a broad iron line is still required with emission extending to within 1.9 gravitational radii of the black hole.  If the iron line comes from an accretion disk truncated at the innermost stable circular orbit, this indicates that the black hole must be spinning rapidly with $a > 0.95$.  Ionized absorption models attempting to explain the $3 - 6 \keV$ spectral curvature without strong gravity predict absorption lines in the $6.4 - 6.6 \keV$ range that are inconsistent with the Chandra spectrum.  The H- and He-like iron absorption lines in the Chandra spectrum are blueshifted by $2.0^{+0.7}_{-0.9} \times 10^3 \kmps$ compared to the source frame, and may originate in a high velocity, high ionization component of the warm absorber outflow.  This high ionization component may dominate the energy budget of the outflow, and account for a significant fraction of the outflowing mass.  Detailed modeling of the warm absorber below 2 keV will be addressed in a later paper, but our results are robust to the broader details of the warm absorber behavior.  The difference spectrum between the high and low flux states is well described by a power law, in agreement with previous studies.

\end{abstract}

% The different journals have different requirements for keywords.  The
% keywords.apj file, found on aas.org in the pubs/aastex-misc directory, 
% contains a list of keywords used with the ApJ and Letters.  These are 
% usually assigned by the editor, but authors may include them in their
% manuscripts if they wish. 


% That's it for the front matter.  On to the main body of the paper.
% We'll only put in tutorial remarks at the beginning of each section
% so you can see entire sections together.

% In the first two sections, you should notice the use of the LaTeX \cite
% command to identify citations.  The citations are tied to the
% reference list via symbolic KEYs.  We have chosen the first three
% characters of the first author's name plus the last two numeral of the
% year of publication.  The corresponding reference has a \bibitem
% command in the reference list below.
%
% Please see the AASTeX manual for a more complete discussion on how to make
% \cite-\bibitem work for you.   

\section{Introduction}

Efficiently accreting black holes are prodigious sources of X-rays.  The illumination of a dense thin accretion disk with X-rays inevitably leads to the production of a strong iron fluorescence line \citep{1989MNRAS.238..729F}.  If the accretion disk lies deep in the potential well of the black hole the fluorescent iron line profile will be broadened and skewed in a characteristic manner by the Doppler and transverse-Doppler shifts of the high velocity orbiting gas and the gravitational redshift and light focusing of the black hole \citep[see, e.g., the review articles by][]{2000PASP..112.1145F, 2003PhR...377..389R}.  As such, the broad iron line is potentially a powerful probe of the astrophysics and spacetime in the immediate vicinity of a black hole.  A major ASCA discovery was just such a line at 6.4 keV from Fe K$\alpha$ in the Seyfert 1 galaxy \mcg6 \citep{1995Natur.375..659T} which showed broadening as large as $\sim 100,000 \kmps$.  Subsequent studies of other AGN have reinforced the existence of broad lines \citep{1995MNRAS.272L...9M, 1997ApJ...477..602N}.

The broad iron line in \mcg6 has been extensively studied with ASCA \citep{1996MNRAS.282.1038I, 1999MNRAS.306L..19I, 2002MNRAS.333..687S}, BeppoSAX \citep{1999A&A...341L..27G}, RXTE \citep{1999MNRAS.310..973L, 2000MNRAS.318..857L, 2001ApJ...548..694V}, Chandra \citep{2002ApJ...570L..47L} and XMM-Newton \citep{2001MNRAS.328L..27W, 2002MNRAS.335L...1F, 2004MNRAS.348.1415V, 2004MNRAS.349.1153R}.  All observations are consistent with a broad highly-redshifted  disk line feature.  Most alternative mechanisms for broadening the iron emission line (i.e. those that do not require strong gravity) are unlikely to work \citep[e.g.,][]{1995MNRAS.277L..11F, 2000ApJ...533..821R, 2000MNRAS.317L..11R}.  We present a deep Chandra High Energy Transmission Grating Spectrometer (HETGS) observation of \mcg6 and, concentrating on the iron K bandpass, describe the narrow emission and absorption features in the spectrum above 2 keV, and constrain the column densities and ionization parameters of highly ionized absorbing gas and discuss the implications.  We investigate whether the broad iron line is not an emission feature but the result of a curved continuum spectrum caused by ionized absorption \citep[e.g.,][]{2003PhDT.........8K}.  Detailed modeling of the warm absorber below 2 keV will be addressed in a later paper, and our results are robust to the broader details of the warm absorber behavior.

\section{Observations}

\begin{figure}
\centerline{\includegraphics[scale=0.35,angle=270]{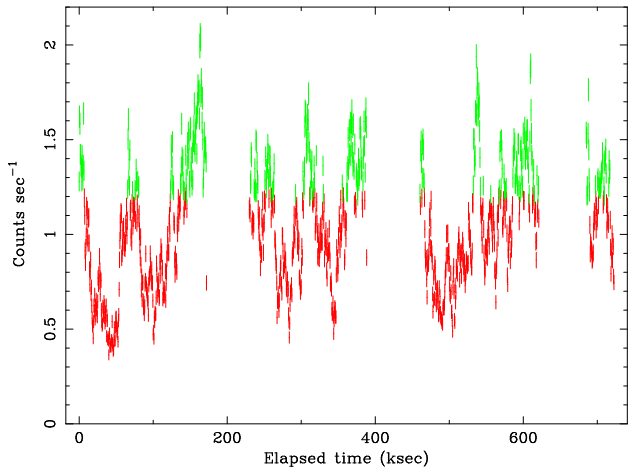}}
\caption{Combined HEG plus MEG $0.45 - 10 \keV$ first order light curve of \mcg6.  The high (green) and low (red) flux states are separated by $1.2$ count $\ps$.  \label{fig:lc}}
\end{figure}

\mcg6 ($z = 0.007749$) was observed by the Chandra HETGS between 2004-05-19 and 2004-05-27 (obs. ids. 4759, 4760, 4761 and 4762) resulting in a good exposure time of 522 ksec.  Spectra and instrument responses were generated using the standard CIAO tools and analyzed using the ISIS spectral fitting software \citep{2002hrxs.confE..17H}.  For improved accuracy, we correct the individual S0 to S5 chip effective areas (ARFs) before combining them to generate the Medium Energy Gratings (MEG) and High Energy Gratings (HEG) +1 and -1 effective areas used for data analysis.  To do so, we correct the quantum efficiencies (QE) of the S0, S2, S4, and S5 front-illuminated (FI)  chips to better agree with the QEs of the back-illuminated (BI) S1 and S3 chips using a  publicly available correction factor \citep{2004SPIE.5165..457M}\footnote{A BI:FI correction file is available on the web at {\tt http://space.mit.edu/ASC/calib/ficorr.txt}.}.  The correction is $< 10\pct$ above 900~eV and $< 19\pct$ below.  Once corrected, the CIAO tool {\it dmarfadd} was used to combine the S0-S5 effective areas to create the -1 and +1 ARFs.  The $+1$ and $-1$ orders of the HEG were combined, as were the $+1$ and $-1$ orders of the MEG.  The HEG and MEG spectra were modeled separately.  The zero order spectrum is heavily piled up ($\approxgt 90\pct$ given an observed zero order rate of $\simeq 0.7$ counts per frame) and is not used.  The combined HEG plus MEG first order $0.45 - 10\keV$ count rate varies between $\simeq 0.4$ and 2 count $\ps$ and the light curve is shown in Fig.~\ref{fig:lc}.  The error bars quoted throughout the paper are the $90\pct$ confidence interval ($\Delta \chi^2 = 2.706$ for one interesting parameter), unless explicitly stated otherwise.

\subsection{Broad Iron Line} \label{sec:bl}

\begin{figure}
\centerline{\includegraphics[scale=0.35,angle=270]{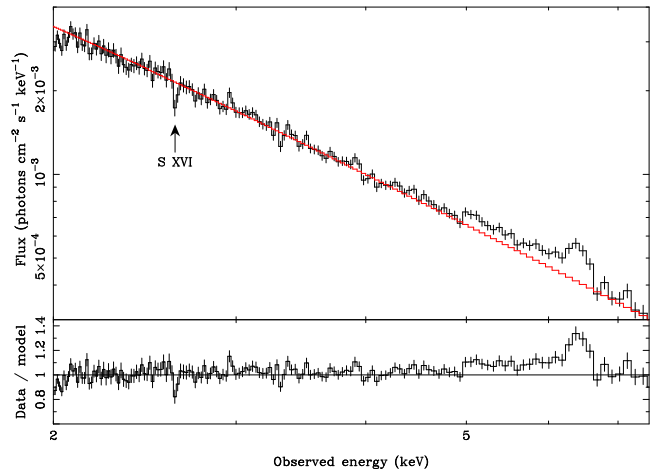}}
\caption{Heavily binned HEG spectrum of \mcg6 (upper panel) with a power law joining the $2.5 - 3\keV$ and $8 - 9\keV$ bands overlaid (red line), and the ratio of the data to power law model (lower panel).  The heavy binning reveals the broad iron K$\alpha$ line that peaks around 6.4 keV and is extremely redshifted down to $\approxlt 5 \keV$.  The arrow indicates \ion{S}{16} Ly$\alpha$ absorption at 2.62~keV.  \label{fig:broad_fe}}
\end{figure}

To show the broad iron line the high resolution HEG spectrum was heavily binned to 0.067 \AA\ bin$^{-1}$ and an absorbed power law model was used to join the $2.25 - 2.5 \keV$ and $7 - 7.5 \keV$ bands (see Fig.~\ref{fig:broad_fe}).  The best fit photon index is $\Gamma = 1.82^{+0.05}_{-0.04}$ with the neutral absorbing column density fixed at $N_H = 2.1 \times 10^{21} \pcmsq$ to approximate the effects of the warm absorber \citep[a few per cent above $\simeq 3 \keV$;][]{2001ApJ...554L..13L}.

\begin{figure}
\centerline{\includegraphics[scale=0.35,angle=270]{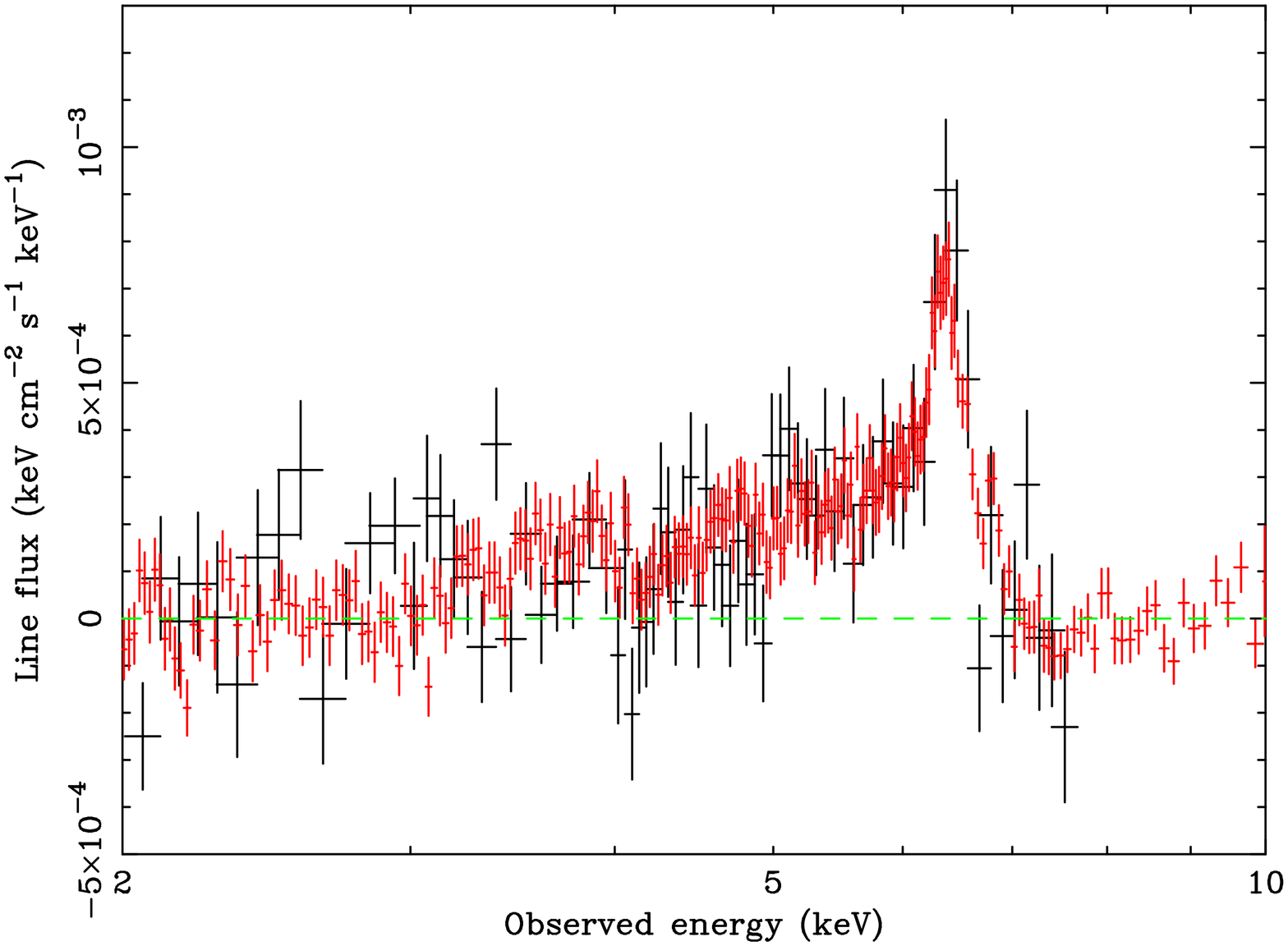}}
\caption{\mcg6 broad iron line from Chandra HEG (black points) and XMM-Newton EPIC pn \citep[][red points]{2004MNRAS.348.1415V} shown in flux units.  The Chandra data have been heavily binned (0.03 \AA\ bin$^{-1}$ above 4 keV and more heavily below 4 keV).  This plot was created by taking the ratio of each data set to the appropriate underlying power law continuum model (see Section~\ref{sec:bl}) and multiplying by the power law model normalization.  There is excellent agreement between the broad iron line profiles even though the Chandra and XMM-Newton observations were not contemporaneous.  The ``red wing'' of the iron K line extends from $\simeq 6 \keV$ down to $\approxlt 3 - 5 \keV$, and the ``narrow core'' peaks around 6.4~keV.  \label{fig:chandra_xmm}}
\end{figure}

The broad iron line profile is very similar to that seen in previous observations and a comparison between the Chandra and XMM-Newton \citep{2004MNRAS.348.1415V} line profile is shown in Fig.~\ref{fig:chandra_xmm}.  The ``red wing'' of the iron K line extends from $\simeq 6 \keV$ down to $\approxlt 3 - 5 \keV$, and the ``narrow core'' peaks around 6.4~keV.  The XMM-Newton spectrum has a slightly different continuum model, namely an absorbed power law joining the $2 - 3 \keV$ and $8.5 - 10 \keV$ bands with neutral $N_H = 3.6 \times 10^{21} \pcmsq$ and $\Gamma = 1.98$.  There is remarkable agreement between the Chandra and XMM-Newton spectra even though they were not contemporaneous.

The continuum and broad iron line in the Chandra HEG spectrum are well described ($\chi^2 / {\rm d.o.f.} = 144 / 123$) by a relativistic disk line plus reflection model.  The reflection spectrum is modeled by {\tt pexrav} \citep{1995MNRAS.273..837M} plus a narrow Gaussian fixed at 6.4~keV in the source-frame.  The {\tt pexrav} model spectrum consists of two components, a power law plus a ``reflected'' or back-scattered continuum that is produced by a power law spectrum illuminating a slab of neutral gas.  The {\tt pexrav} model does not include the iron fluorescence line and hence that is added as a narrow Gaussian at 6.4~keV.  The reflection fraction (the ratio of the normalization of the reflected spectrum to the normalization of the power law, with a value of 1 for an isotropic source above the disk) is fixed at 2.2 and the iron abundance is fixed at three times the Solar value based on previous XMM-Newton models \citep{2002MNRAS.335L...1F}.  Both the reflection and iron line component are blurred by a disk line kernel for emission around a spinning Kerr black hole \citep{1991ApJ...376...90L}, and everything is absorbed by a fixed neutral column density.  In addition a narrow, neutral Fe K$\alpha$ line is added at 6.4~keV in the source frame to represent reflection off distant material, e.g., the narrow line region or torus.  The model parameters are given in Table~\ref{tab:broad_fe}.

The HETGS spectrum is best suited to studying narrow spectral features and does not strongly constrain the broad line properties.  A model in which the reflection fraction is fixed at $R = 1$ and the iron abundance fixed at the Solar value also provides a good description of the spectrum with similar broad iron line parameters and an almost identical $\chi^2 / {\rm d.o.f.}$ of $143 /123$.  Since we cannot easily distinguish between such broadband models we have adopted the values used by \citet{2002MNRAS.335L...1F} (i.e. $R = 2.2$ and an iron abundance of three times the Solar value), and the line energies and equivalent widths we measure are insensitive to this choice.  We can, however, show that the HEG spectrum is better described by a neutral broad iron line than an ionized broad iron line.  If the iron line added to the reflection spectrum is at 6.7~keV instead of 6.4~keV the best fit model has a worse $\chi^2 / {\rm d.o.f.} = 152 / 123$.  We do not include the weak reflection continuum associated with the narrow 6.4~keV line, and this does not affect our results.

\begin{deluxetable*}{cccccccccc}
\tablecaption{Continuum and iron line parameters\tablenotemark{a}  \label{tab:broad_fe}}
\tablecolumns{10}
\tablehead{\colhead{$N_H$\tablenotemark{b}} & \colhead{$\Gamma$} & \colhead{$K$\tablenotemark{c}} & \colhead{$r_{\rm in}$} & \colhead{$r_{\rm out}$\tablenotemark{b}} & \colhead{$q$\tablenotemark{d}} & \colhead{$i$} & \colhead{EW (broad)} & \colhead{EW (narrow)} & \colhead{$\chi^2$ / dof} \\
\colhead{($\pcmsq$)} & \colhead{} & \colhead{} & \colhead{($r_g$)} & \colhead{($r_g$)} & \colhead{} & \colhead{($\degmark$)} & \colhead{(eV)} & \colhead{(eV)} }
\startdata

$2.1 \times 10^{21}$ & $1.94^{+0.01}_{-0.01}$ & $1.6 \times 10^{-2}$ & $1.3^{+12.5}_{-0.1}$ & 200 & $2.3^{+2.5}_{-0.1}$ & $26^{+3}_{-4}$ & $134^{+64}_{-11}$ & $15^{+10}_{-9}$ & 144 / 123 \\

\enddata
\tablenotetext{a}{The reflection fraction $R$ is fixed at 2.2 and the iron abundance ${\rm A}_{\rm Fe}$ is fixed at three times the Solar value based on previous XMM-Newton observations \citep{2002MNRAS.335L...1F}.  The broad Fe K$\alpha$ line energy is fixed at a source-frame energy of $6.40 \keV$.}
\tablenotetext{b}{Parameter fixed.}
\tablenotetext{c}{The normalization of the {\tt pexrav} model $K = \phpcmsqpspkeV$ of the power law component at 1 keV in the observed frame.}
\tablenotetext{d}{The radial emissivity profile of the disk line is $\propto r^{-q}$.}
\end{deluxetable*}

\subsection{Narrow Iron Line Components}

\begin{figure}
\centerline{\includegraphics[scale=0.35,angle=270]{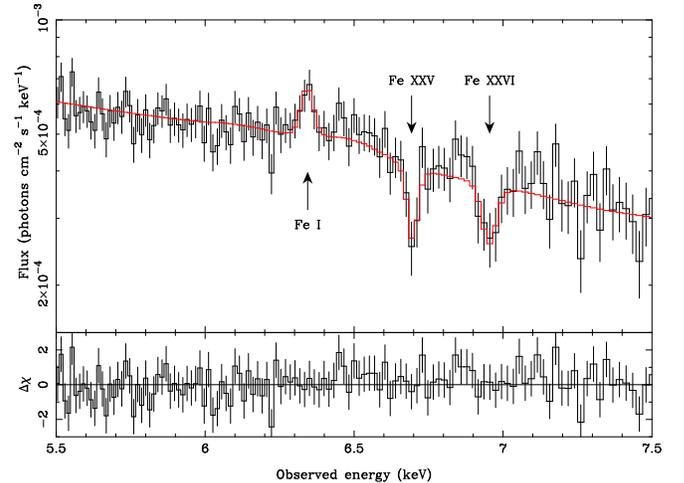}}
\caption{\mcg6 Chandra HEG spectrum of the iron K bandpass for the entire observation.  The HEG $\pm 1$ orders have been combined and there are 0.0056 \AA\ bin$^{-1}$.  The upper panel shows the data (black points) and best fitting model folded through the instrument response (red line), and the lower panel shows the $\chi$ residuals to the fit.  Note the narrow Fe K$\alpha$ emission at 6.34~keV and the narrow absorption features at 6.69~keV and 6.96~keV.  The underlying continuum model consists of a reflection component plus a neutral Fe K$\alpha$ line, and both components have been blurred using a relativistic disk line kernel.  \label{fig:heg_avg}}
\end{figure}

The unbinned HEG spectrum (0.0056 \AA\ bin$^{-1}$; the FWHM of the HEG is 0.012 \AA) shows a number of narrow emission and absorption features in the iron K bandpass (Fig.~\ref{fig:heg_avg}).  The $5.5 - 7.5 \keV$ continuum and broad iron line were modeled as described in Section~\ref{sec:bl} and Table~\ref{tab:broad_fe}, with most of the parameters fixed except the normalizations of the continuum and broad line and the parameters of the Gaussian representing the neutral Fe K$\alpha$ line.  Individual absorption lines were added as Gaussians with their centroids, strengths and widths being free parameters.  The best fit line parameters are given in Table~\ref{tab:narrow_fe}.

There is a narrow emission line at $6.39 \keV$ (source frame), consistent with neutral iron K$\alpha$ fluorescence, that has an EW of $18 \eV$ \citep[in good agreement with][]{2002ApJ...570L..47L} and a FWHM of $< 4700 \kmps$.  We also detect two unresolved absorption features, one at $6.69 \keV$ (observed) consistent with the 1s -- 2p resonance line of He-like iron \citep[6.70 keV;][]{1996ADNDT..64....1V} and the second at $6.96 \keV$ (observed) consistent with the two 1s -- 2p resonance lines of H-like iron \citep[6.95 keV and 6.97 keV;][]{1996ADNDT..64....1V}.  Both of these absorption features are consistent with zero redshift or, if associated with \mcg6, an outflow velocity of $2.0^{+0.7}_{-0.9} \times 10^3 \kmps$ for \ion{Fe}{25} and $1.9^{+0.7}_{-0.8} \times 10^3 \kmps$ for \ion{Fe}{26}.

\begin{figure}
\centerline{\includegraphics[scale=0.35,angle=270]{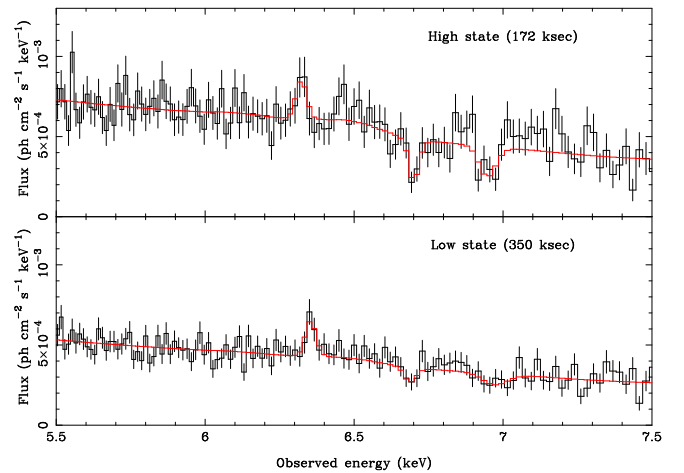}}
\caption{\mcg6 HEG spectra of the iron K bandpass for the high flux state (top panel) and low flux state (bottom panel).  \label{fig:hilo_fe}}
\end{figure}

\subsection{Variability} \label{sec:var}

\begin{figure}
\centerline{\includegraphics[scale=0.35,angle=270]{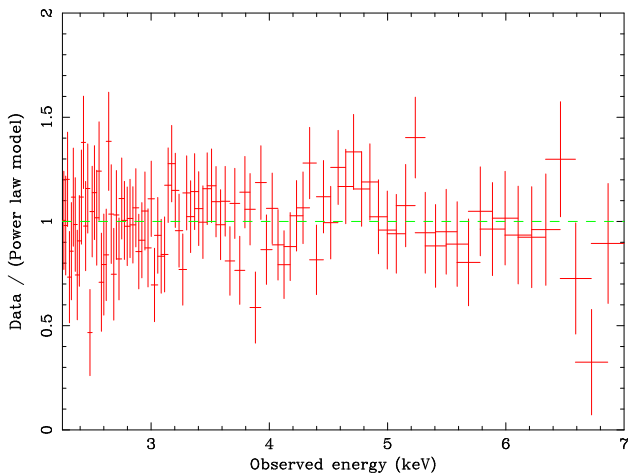}}
\caption{Ratio of the difference between the high and low flux state spectra to a power law model with photon index $\Gamma = 2$.  Note the suggestion of an absorption line at 6.7 keV. \label{fig:hilo_pl}}
\end{figure}

We now compare the spectrum of \mcg6 in a high and low flux state since this can help determine whether certain spectral features are caused by absorption or emission.  Using the combined first order HEG plus MEG light curve of \mcg6 with 500~s bins (Fig.~\ref{fig:lc}) we extracted high and low flux state spectra from those bins with count rates greater than or less than $1.2$~counts~$\ps$, respectively.  The effective exposure times are 172 ksec for the high flux state and 350 ksec for the low flux state, chosen so that the signal-to-noise ratio of each spectrum is comparable.  The high and low flux state HEG spectra are shown in Fig.~\ref{fig:hilo_fe} and the parameters of the narrow iron emission and absorption lines are given in Table~\ref{tab:narrow_fe}.  The line centroid energies are consistent between the high and low flux states.  The difference in the continuum spectrum between the high and low flux states in the 2.25 -- 7 keV band is well described ($\chi^2_\nu = 0.9$) by a power law of photon index $\Gamma = 2.0^{+0.2}_{-0.1}$ (see Fig.~\ref{fig:hilo_pl}), in agreement with previous findings that the variability is dominated by a power law component \citep[e.g.,][]{2002MNRAS.335L...1F, 2003MNRAS.346..833T, 2003MNRAS.340L..28F, 2004MNRAS.348.1415V}.  Furthermore, there is no evidence of continuum curvature or systematic deviation from a power law model in the 2.25 -- 7 keV band.  A narrow He-like iron absorption feature is weakly detected at approximately 6.7 keV in the difference spectrum.

\begin{deluxetable*}{cccccc}
\tablecaption{Narrow emission and absorption features \label{tab:narrow_fe}}
\tablecolumns{6}
\tablehead{\colhead{Flux state} & \colhead{Identification} & \colhead{Laboratory energy\tablenotemark{a}} & \colhead{Line energy (keV)} & \colhead{FWHM} & \colhead{EW} \\
\colhead{} & \colhead{} & \colhead{(keV)} & \colhead{Source frame (observed)\tablenotemark{b}} & \colhead{(1,000 \kmps)} & \colhead{(eV)}}
\startdata

Average & \ion{Si}{14} & 2.01 & $2.005(1.990)^{+0.001}_{-0.005}$ & $< 0.7$ & $-2^{+0}_{-1}$ \\

& \ion{S}{16} & 2.62 & $2.640(2.620)^{+0.006}_{-0.004}$ & $< 3.7$ & $-5^{+2}_{-2}$ \\

& \ion{Fe}{1} -- \ion{Fe}{19} & 6.40 -- 6.47 & $6.393(6.344)^{+0.106}_{-0.014}$ & $<4.7$\tablenotemark{c} & $18^{+11}_{-8}$ \\

& \ion{Fe}{25} & 6.70 &  $6.745(6.693)^{+0.018}_{-0.006}$ & $<2.0$ & $-18^{+7}_{-5}$ \\

& \ion{Fe}{26} & 6.95, 6.97 & $7.009(6.955)^{+0.018}_{-0.017}$ & $<3.8$ & $-21^{+10}_{-11}$ \\

Low & \ion{Fe}{1} -- \ion{Fe}{14} & 6.40 -- 6.41 & $6.407(6.367)^{+0.008}_{-0.016}$ & $< 2.7$ & $25^{+10}_{-9}$ \\

& \ion{Fe}{25} & 6.70 & $6.745(6.693)^{+0.017}_{-0.021}$ & $< 3.2$ & $-13^{+9}_{-9}$ \\

& \ion{Fe}{26} & 6.95, 6.97 & $7.028(6.974)^{+0.049}_{-0.042}$ & $< 9.0$ & $-23^{+15}_{-20}$ \\

High & \ion{Fe}{1} & 6.40 & $6.372(6.323)^{+0.017}_{-0.017}$ & $< 2.6$ & $19^{+11}_{-11}$ \\

& \ion{Fe}{25} & 6.70 & $6.762(6.710)^{+0.001}_{-0.019}$ & $< 2.3$ & $-25^{+9}_{-9}$ \\

& \ion{Fe}{26} & 6.95, 6.97 & $7.003(6.949)^{+0.018}_{-0.017}$ & $2.0^{+1.8}_{-1.7}$ & $-32^{+14}_{-16}$ \\

\enddata
\tablenotetext{a}{Values for He- and H-like ions from \citet{1996ADNDT..64....1V} and for \ion{Fe}{1} -- \ion{Fe}{19} from \citet{1969ApJS...18...21H}.}
\tablenotetext{b}{The first figure gives the line energy in the source frame and the second figure (in parentheses) gives the observed line energy.}
\tablenotetext{c}{In this case the line energy was fixed at the best fit value.  If the line energy is a free parameter there is an ambiguity distinguishing between the broad and narrow line components and the FWHM is only weakly constrained to be $< 21.8 \times 10^3 \kmps$.}
\end{deluxetable*}

\section{Discussion}

\subsection{Column Densities} \label{sec:col_dens}

We can estimate the column densities of \ion{Fe}{25} and \ion{Fe}{26} ions from the equivalent widths of their absorption lines \citep[see, e.g., ][]{1978ppim.book.....S, 2000ApJ...539..413K}.  The inferred column density depends on the absorption line profile and we assume the absorber has a Maxwellian velocity dispersion with a velocity spread parameter $b = \sqrt{2} \sigma$ where $\sigma$ is the radial velocity dispersion.  The thermal broadening is roughly $b \simeq 20 \kmps$ for reasonable temperatures so larger values of $b$ will imply some bulk motion or turbulence in the absorber(s).  The column of \ion{Fe}{25} required to produce an 18 eV EW absorption line is $N_{\rm Fe XXV} \simeq 3 \times 10^{17} - 3 \times 10^{18} \pcmsq$ for $b = 500 - 100 \kmps$, respectively.  Similarly, the column density of \ion{Fe}{26} required to produce a 21 eV EW absorption line is $N_{\rm Fe XXVI} \simeq 6 \times 10^{17} - 4 \times 10^{19} \pcmsq$ for $b = 500 - 100 \kmps$, respectively.  The corresponding hydrogen column density depends on the iron abundance and fraction of iron in each ionization state.  For \ion{Fe}{25} we estimate $N_{\rm H} \simeq \left(\frac{0.5}{X_{25}}\right) \left(\frac{n_{{\rm H}/{\rm Fe}}}{3 \times 10^4}\right) 2 \times 10^{22} - 2 \times 10^{23} \pcmsq$ for $b = 500 - 100 \kmps$, respectively, where $n_{{\rm H}/{\rm Fe}}$ is the ratio of hydrogen ions to all iron ions ($3 \times 10^4$ for Solar abundance) and $X_{25}$ is the fraction of iron ions found in \ion{Fe}{25} (which peaks around 0.5).  Similarly, for \ion{Fe}{26} we infer $N_{\rm H} \simeq \left(\frac{0.5}{X_{26}}\right) \left(\frac{n_{{\rm H}/{\rm Fe}}}{3 \times 10^4}\right) 4 \times 10^{22} - 3 \times 10^{24} \pcmsq$ for $b = 500 - 100 \kmps$, respectively.  If the H- and He-like Fe absorption lines are produced in a single absorber with $\log \xi \simeq 3.5 - 4$ then $X_{25} \simeq X_{26} \simeq 0.4 - 0.5$ \citep[see Fig.~8 of][]{2001ApJS..133..221K}.

\begin{figure}
\centerline{
    \includegraphics[scale=0.25,angle=270]{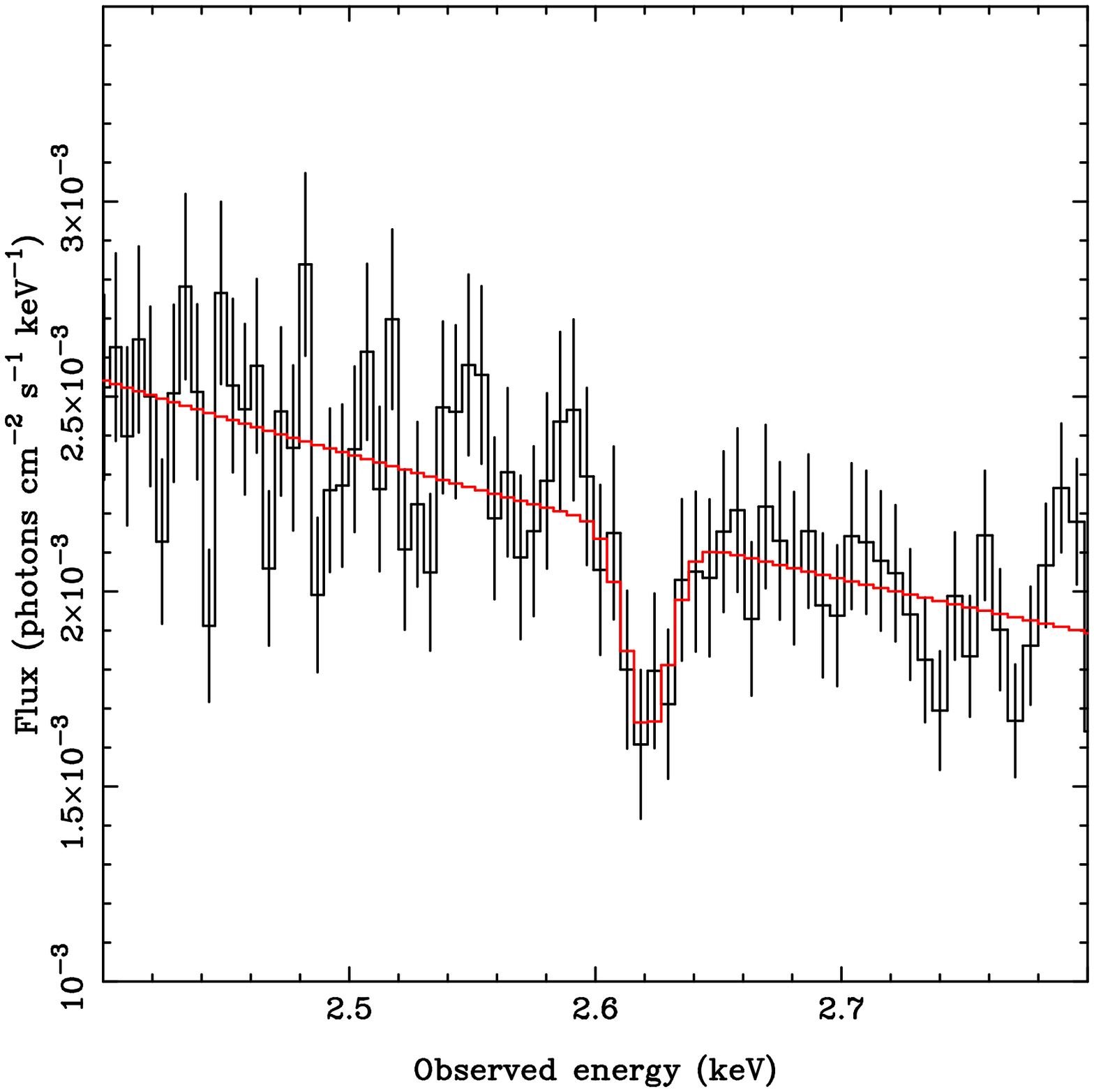}
    \includegraphics[scale=0.25,angle=270]{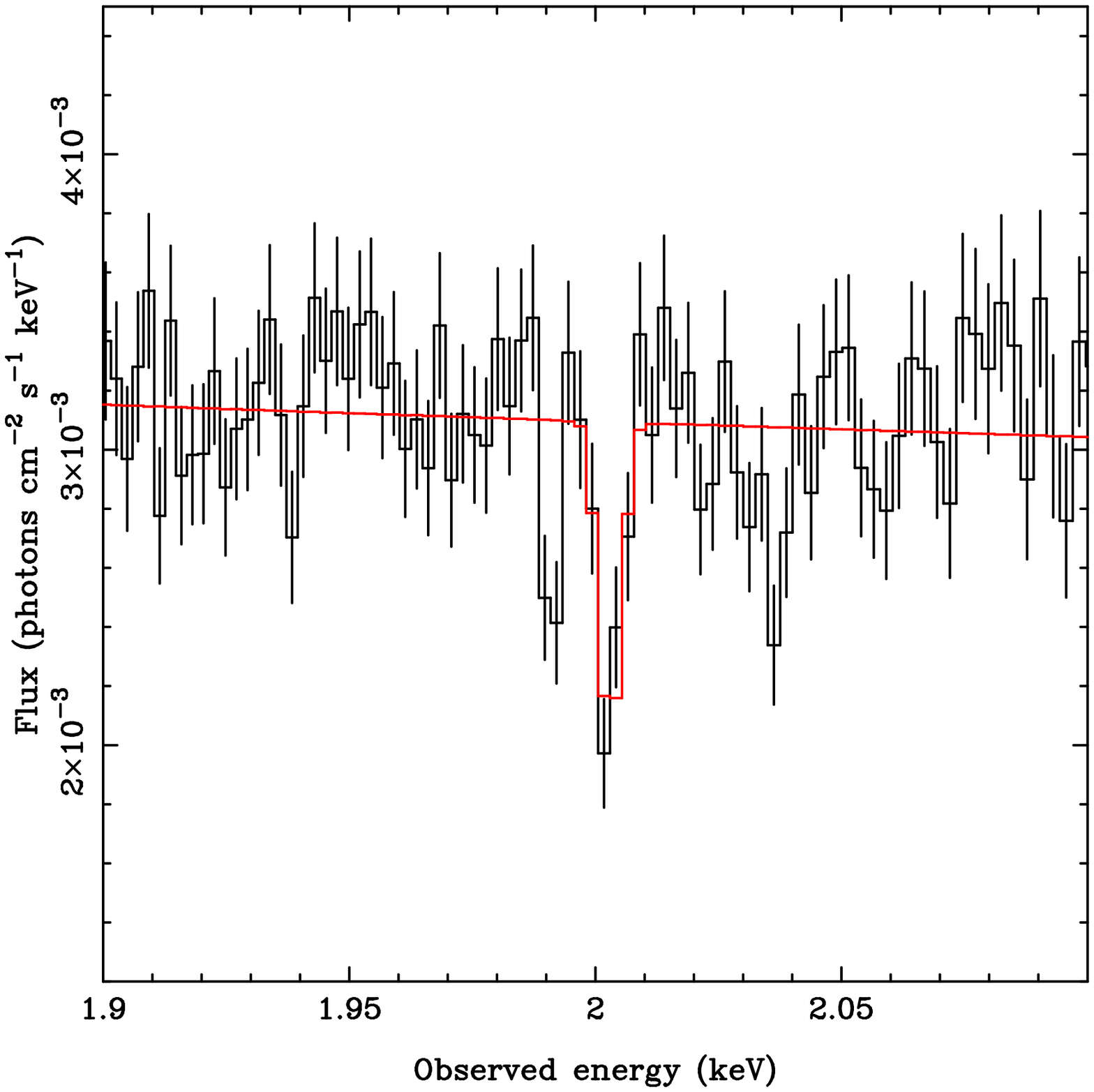}
}
\caption{Chandra HEG spectra showing absorption lines.  \emph{Left panel:} \ion{S}{16} Ly$\alpha$ absorption (2.62~keV in laboratory frame) outflowing from \mcg6 at $\simeq 2200 \kmps$ (red line).  \emph{Right panel:} \ion{Si}{14} Ly$\alpha$ absorption (2.01~keV in laboratory frame) outflowing from \mcg6 at $\simeq 2300 \kmps$ (red line).  The absorption feature observed at 1.99 keV is from \ion{Si}{14} Ly$\alpha$ at rest in the source frame, with an EW of 1 eV.  \label{fig:s_si}}
\end{figure}

If we model the absorbing gas as a photo-ionized plasma illuminated by a power law using the {\tt xstar} code \citep{2001ApJS..133..221K}, with Solar iron abundance and $b = 100 \kmps$, a good fit ($\chi^2_\nu = 1.1$) to the $5.5 - 7.5 \keV$ spectrum is given by a power law plus disk line plus narrow iron K$\alpha$ emission line absorbed by a column density $\log N_H = 23.2^{+0.3}_{-0.6}$ with an ionization parameter $\log \xi = 3.6^{+0.1}_{-0.2}$, where $\xi = L_{\rm ion}/(n_eR^2)$, $L_{\rm ion}$ is the $1 - 1000$~Ry luminosity and $R$ is the distance between the continuum source and cloud.  For three times Solar abundance the column density drops to $\log N_H = 22.9^{+0.1}_{-1.0}$.  The redshift of the absorbing gas is $z = 1.3^{+1.3}_{-1.1} \times 10^{-3}$.  Note also that there is some degeneracy between the ionization parameter and $b$ since both can affect the ratio of \ion{Fe}{25} EW to \ion{Fe}{26} EW.  The best fitting {\tt xstar} model predicts other absorption features that are consistent with the Chandra HEG spectrum.  The strongest lines predicted by the model in the energy range $2 \approxlt E \approxlt 6 \keV$ are \ion{S}{16} Ly$\alpha$ at 2.62~keV and \ion{Si}{14} Ly$\alpha$ at 2.01~keV.  Both of these absorption lines are detected in the HEG spectrum (Fig.~\ref{fig:s_si}, Table~\ref{tab:narrow_fe}), with outflow velocities of $2288^{+687}_{-457} \kmps$ for \ion{S}{16} and $2247^{+80}_{-726} \kmps$ for \ion{Si}{14}.  We have not considered in our fit the spectrum below 2 keV or the warm absorber which will be the subject of a future study.  The column densities in the high and low flux states are statistically consistent with a single value equal to that of the time-averaged spectrum.

\subsection{Robustness of the Broad Iron Line} \label{sec:robust}

It has been proposed that the extremely redshifted ``red wing'' of the iron K line might actually be caused by absorption \citep[e.g.,][]{2003PhDT.........8K}.  In this model the continuum passes through a large column density of moderately ionized gas which causes significant curvature of the transmitted continuum above the iron L edge at $\sim 0.7 \keV$ that extends up to the iron K bandpass ($\approxgt 5 \keV$), and this curvature can mimic the red wing of the putative broad iron K$\alpha$ emission line.  For iron to have L-shell electrons the absorbing gas cannot be as highly ionized as the gas giving rise to the H- and He-like iron absorption lines.  On the other hand, the gas must be sufficiently highly ionized to avoid excessive opacity of soft X-rays, therefore $\xi \approxgt 10 - 100$ \citep{1984ApJ...286..366K}.  The fact that the difference spectrum between the high and low flux states above 2.25 keV (where the effect of the warm absorber is small) is well described by an unabsorbed power law (Section \ref{sec:var}) strongly suggests that the curvature is not an absorption artifact.  If the curvature in the time-averaged hard X-ray spectrum is caused by an intervening absorber with constant optical depth, $\tau = N \sigma (E)$, then the absorption should affect the intrinsic high state, $F_{h}$, and intrinsic low state, $F_{l}$, similarly (i.e. as $e^{-\tau} F_h$ and $e^{-\tau} F_l$, respectively), and hence the difference spectrum should also show the same absorption (i.e. $e^{-\tau} [F_h - F_l]$).  We do not expect $\tau$ to vary between the high and low flux states since the continuum variability timescale is so short (typically hundreds of seconds).  Even though the difference spectrum does not show such an absorption signature above 2.25 keV, in this section we present a comparison between an absorption model and the Chandra HEG spectrum.

\begin{figure}
\centerline{\includegraphics[scale=0.35,angle=270]{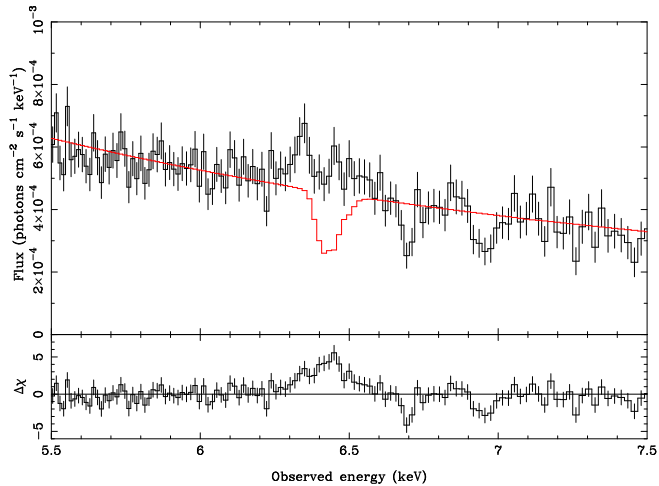}}
\caption{Chandra HEG spectrum (black line) with an ionized absorption model (folded through the instrument response) of the broad red wing of the iron line overlaid (red line).  The ionized absorber model was fit to the $3 - 10 \keV$ XMM-Newton spectrum excluding the $6 - 8 \keV$ range in which the line has a concave profile and narrow core.  The ionized absorber produces a curved continuum spectrum but also predicts iron absorption lines in the $6.4 - 6.6 \keV$ range in the source frame that are inconsistent with the Chandra HEG spectrum.  We conclude that a simple absorption model cannot account for the broad red wing of the iron line.  \label{fig:xi_abs}}
\end{figure}

To test the ionized absorption hypothesis, we modeled the $3 - 10 \keV$ XMM-Newton EPIC-pn spectrum \citep{2004MNRAS.348.1415V} with an {\tt xstar} model absorbing a power law spectrum, excluding the $6 - 8 \keV$ range in which the line has a more complicated profile and narrow core (see Fig.~\ref{fig:chandra_xmm}).  The curvature of the broad red wing is approximately described by a model with Solar iron abundance, $b = 100 \kmps$, $\log N_H = 22.6^{+0.1}_{-0.0}$, $\log \xi = 2.2^{+0.1}_{-0.1}$ and $\Gamma = 2.12^{+0.03}_{-0.02}$.  The XMM-Newton spectrum was used to construct this model since it provides better statistical constraints on very broad spectral features than the Chandra HEG spectrum.  The ionized absorption model predicts a complex of narrow absorption lines between approximately 6.4 and 6.6 keV in the source frame that are inconsistent with the Chandra HEG spectrum, as shown in Fig.~\ref{fig:xi_abs}.  Narrow $1s-2p$ iron absorption lines \citep[from \ion{Fe}{18} at 6.4 keV up to \ion{Fe}{23} at 6.6 keV;][]{2002ApJ...570..165B} are a generic feature of ionized absorption models in which there is significant iron L-shell absorption.  We have confirmed that for $b = 0 \kmps$ a similarly strong absorption line complex is produced that would be readily visible in the HEG spectrum, and the strength of the absorption lines will increase with increasing $b$.  We conclude that a simple absorption model cannot account for the broad red wing of the iron line.

More complex absorption models are harder to constrain.  It is possible that emission lines fill in the absorption lines, although this requires fine tuning the ionization state and strength of emission so that no significant emission or absorption residuals are seen between $\sim 6.4$ and 6.5 keV, and we consider this scenario unlikely.  A neutral partial covering model can be ruled out because it is inconsistent with the RXTE spectrum above 10~keV \citep{2004MNRAS.349.1153R}.  An ionized partial covering model \citep[e.g., see][]{2004astro.ph..9091T} cannot simply be ruled out, but if the partial covering component produces a strong absorption line similar to the simple absorption model discussed above we can constrain the covering fraction to be $f \approxlt 5\pct$.  For a larger covering fraction the absorption features should be visible in the HEG spectrum.  It is difficult to introduce sufficient continuum curvature to account for the red wing of the broad iron line with such a small covering fraction.  It is also important to reiterate that the spectral variability of \mcg6 is inconsistent with an absorption model \cite[Section \ref{sec:var}, Section \ref{sec:robust} and][]{2002MNRAS.335L...1F}.

The presence of the highly ionized absorber giving rise to the H- and He-like iron absorption lines seen in the Chandra HEG spectrum at 7.0 keV and 6.7 keV, respectively, does not significantly affect the modeling of the red wing of the iron K line $\approxlt 6 \keV$ and hence the conclusion that the black hole must be spinning rapidly is robust.  An excellent fit ($\chi^2_\nu = 0.9$) to the $3 - 10 \keV$ XMM-Newton EPIC-pn spectrum is given by a power law absorbed by the highly ionized {\tt xstar} model used to describe the Chandra HEG spectrum (i.e. Solar abundance, $b = 100 \kmps$, $\log N_H = 23.2$, $\log \xi = 3.6$) plus a Laor disk-line and narrow Fe $K\alpha$ emission line.  The highly ionized absorber does not introduce sufficient curvature to the continuum to significantly affect the red wing of the iron line, and we find that the inner radius of the Laor disk line model is constrained to be $r_{\rm in} < 1.9 r_g$ with $99.9\pct$ confidence (for one interesting parameter) indicating that the broad red wing of the iron line is still present.  If the Fe line comes from an accretion disk truncated at the innermost stable circular orbit, this indicates a black hole spin of $a > 0.95$.  In this model the radial emissivity profile is $r^{-q}$ with $q = 3.4^{+0.1}_{-0.2}$, the disk inclination angle is $i = 42 \pm 1\degmark$ and the EW of the narrow Fe K$\alpha$ line is 22 eV, consistent with the Chandra value.

The lower inclination angle of $i = 26$ given in Table~\ref{tab:broad_fe} results from attributing the drop in flux at 6.69~keV to the blue wing of the broad iron line rather than an absorption line.  With the CCD resolution XMM-Newton data the $i = 26\degmark$ interpretation wants an additional emission line at 6.9~keV \citep{2002MNRAS.335L...1F} to account for the flux above $\approxgt 6.7 \keV$.  The value of $i = 42\degmark$ in our model is consistent with, e.g., Model 2 of \citet{2002MNRAS.335L...1F}.  Furthermore, the $i = 42\degmark$ model results in extra broad iron line flux up to $\simeq 6.9\keV$ and can account for the positive residuals around $6.8 - 6.9 \keV$ seen, e.g., in Fig.~\ref{fig:heg_avg} which assumed an inclination of $i = 26\degmark$.

\subsection{The High Ionization Absorber}

The highly ionized gas giving rise to the H- and He-like absorption lines may be either intrinsic to the source or serendipitously located along the line of sight.  In the former case, it must be outflowing from \mcg6 at $2.0^{+0.7}_{-0.9} \times 10^3 \kmps$, whereas in the latter case it would have a velocity $0.3^{+0.7}_{-0.9} \times 10^3 \kmps$ relative to the Galaxy.

The possibility that the absorbing gas is local is intriguing since a number of other AGN show iron absorption features that are kinematically consistent with a local origin \citep{2004astro.ph..8506M}.  In addition, \ion{O}{7} absorption is seen along every line of sight through the Galaxy for which we have a sufficiently high signal-to-noise ratio spectrum \citep{2003ApJ...586L..49F, fang_abs}, although the O absorber need not be as highly ionized as the Fe absorber.  If the H- and He-like Fe absorption seen in \mcg6 is simply due to a high ionization phase of the inter-stellar medium (ISM) then we can estimate the size of the absorber.  If we assume the absorber is in pressure balance with the ISM, $P/k = 10^4 \pcmcu$ \citep{1987ARA&A..25..303C}, and is collisionally ionized, $T \sim 10^7 \K$, then using $P/k = nT$ implies $n = 10^{-3} \pcmcu$.  The path length through the absorber, $l$, is then $l \simeq N_H / n = 30 \Mpc$ for $N_H = 10^{23} \pcmsq$, which is more than three orders of magnitude larger than the diameter of the Galaxy.  If the absorber is local then such a large column density of highly ionized material along a ``random'' line of sight is hard to explain.

An alternative origin for the H- and He-like absorption lines is in an outflow.  \mcg6 is known to possess a complex ``warm absorber'' \citep{1996PASJ...48..211O, 1997MNRAS.286..513R, 1998ApJS..114...73G, 2001ApJ...554L..13L, 2003MNRAS.346..833T, 2004MNRAS.353..319T} showing a broad range of ionization states and outflow velocities, and it is possible that the high ionization absorber is at the hot and fast end of a continuum of properties within the warm absorber \citep{2002xsac.conf....9L}.  Warm absorber models do show components outflowing from \mcg6 at $\simeq 2000 \kmps$ \citep{2003MNRAS.346..833T, 2003ApJ...596..114S}, and discrete, blueshifted absorption lines are seen, such as \ion{O}{7} Ly$\alpha$ and \ion{O}{8} Ly$\alpha$ \citep[outflowing at $2300 \kmps$ and $1900 \kmps$, respectively, consistent with][]{2004MNRAS.353..319T}, \ion{S}{16} Ly$\alpha$ and \ion{Si}{14} Ly$\alpha$ ($2300 \kmps$ and $2200 \kmps$, respectively).  Furthermore, absorption lines are seen in the HEG outflowing at intermediate velocities, such as \ion{Fe}{18} (or \ion{Ne}{5}) at $347 \kmps$ and \ion{Fe}{24} (or \ion{Ne}{9}) at $864 \kmps$ \citep[these are also seen with XMM-Newton;][]{2004MNRAS.353..319T}.  The presence of these high and intermediate velocity components supports the suggestion that the highly ionized absorber is associated with the warm absorber.  An example of a similarly highly ionized, but higher velocity, outflow is the quasar MR~2251-178 for which the HETGS also reveals \ion{Fe}{26} absorption \citep{rob_mr2251}.

The warm absorber itself has a complex internal structure since, for example, \ion{O}{7} Ly$\alpha$ and \ion{O}{8} Ly$\alpha$ absorption lines are also seen at rest in the source frame \citep{2004MNRAS.353..319T} indicating a range of outflow velocities.  Furthermore, strong absorption lines are also seen from much lower ionization species such as \ion{O}{1} that cannot be in equilibrium with the highly ionized absorber (which is too highly ionized to contain any significant \ion{O}{1}), indicating a range of different ionization states distributed throughout the absorber.  This complexity is well known, and while the overall ionization and velocity structure of warm absorbers is unknown it is often modeled by a number of independent, discrete ``zones'' that can adequately reproduce the observed spectrum.

We can estimate the mass outflow rate associated with the high ionization component giving rise to the H- and He-like iron absorption lines using the ionization parameter, $\log \xi = 3.6$, ionizing luminosity, $L_{\rm ion} = 2 \times 10^{43} \ergps$ (assuming $H_0 = 70 \kmpspMpc$, $q_0 = 0$), and outflow velocity, $v = 2000 \kmps$.  The ionization parameter $\xi = L_{\rm ion} / nr^2$ implies $nr^2 = 5 \times 10^{39} {\rm cm}^{-1}$.  The mass outflow rate is then $\dot M_{\rm high} = \Omega n r^2 m_p v = 0.3 (\Omega / 4 \pi) \Msunpyr$, giving a kinetic power of $L_{\rm KE,\ high} = 4 \times 10^{41} (\Omega / 4 \pi) \ergps$ which is approximately $0.1 L_{\rm x}$.  We note that these estimates depend on the unknown extent to which the absorbing material is clumped.  \citet{2005A&A...431..111B} find that, for \mcg6, the warm absorber has an outflow rate of $\dot M_{\rm WA} = 0.16 \Msunpyr$ and a kinetic luminosity of $L_{\rm KE,\ WA} = 10^{39} \ergps$.  For a covering fraction of $(\Omega / 4 \pi) = 0.05$, the kinetic luminosity of the high ionization component, $L_{\rm KE,\ high} = 2 \times 10^{40} \ergps$, dominates the energy budget of the outflow as a whole.  The high ionization component can also account for a significant fraction of the outflowing mass.

An estimate of the distance from the ionizing source can be obtained by using $N \simeq n \Delta r$, where $\Delta r$ is the thickness of the absorber, and the ionization parameter $\xi = L_{\rm ion} / nr^2$, to get $r \le 0.02 (\Delta r / r) \pc$ for $N = 10^{23} \pcmsq$ (Section \ref{sec:col_dens}) and $\Delta r / r \le 1$.  This is extremely close to the central engine and implies a density of $n = 2 \times 10^6 \pcmcu$.  The corresponding timescale for variability due to the motion of the absorber is $r / v \sim 8 (\Delta r / r) \yr$.  We have examined the AO1 Chandra HETGS observation of \mcg6 (obs. id. 433), and detect \ion{Fe}{25} absorption at 6.7 keV with an EW of $-16^{+0}_{-8} \eV$ and \ion{Fe}{26} absorption at 7.0 keV with an EW of $-13^{+11}_{-13} \eV$ \citep[these can be seen in Fig.~5 of][]{2002ApJ...570L..47L}, and are consistent with the values listed in Table~\ref{tab:narrow_fe}.  There is no evidence of variability in the EWs of the H- and He-like Fe lines between the AO1 observations of 2000-04-05 and 2000-08-21--22, and our more recent observations between 2004-05-19 and 2004-05-27, although our measurement uncertainties do not provide strong constraints.

\subsection{Conclusions}

The Chandra HEG spectrum of the iron K bandpass of \mcg6 shows the following.

1.  There are narrow absorption lines in the Chandra HEG spectrum from H- and He-like iron, requiring ion column densities of $N_{\rm Fe XXV} = 3 \times 10^{17} - 3 \times 10^{18} \pcmsq$ for $b = 500 - 100 \kmps$, respectively, and $N_{\rm Fe XXVI} = 6 \times 10^{17} - 4 \times 10^{19} \pcmsq$ for $b = 500 - 100 \kmps$, respectively.  If the absorbing gas is photoionized with Solar iron abundance and a velocity dispersion parameter of $b = 100 \kmps$ we find $\log N_H = 23.2$ and $\log \xi = 3.6$.  The strongest two absorption lines predicted by the model between 2 and 6~keV are also detected, namely \ion{S}{16} Ly$\alpha$ and \ion{Si}{14} Ly$\alpha$.

2.  The difference spectrum between the high and low flux states is well described by a power law in which a 6.7 keV absorption line is weakly detected, showing that the variable power law emission also passes through the highly ionized absorber giving rise to the narrow H- and He-like absorption lines in the average spectrum.

3.  Ionized absorption models, in which continuum curvature can mimic the broad red wing of the putative broad iron K line, predict iron K-shell absorption lines in the 6.4 -- 6.6 keV range that are inconsistent with the Chandra HEG spectrum.  In addition, the fact that the difference spectrum between the high and low flux states is well described by a power law strongly suggests that the curvature is not caused by absorption.

4.  Applying the photoionization model used to described the H- and He-like absorption lines in the Chandra HEG spectrum to the XMM-Newton EPIC-pn spectrum we find that a broad iron line is still required.  The Fe K$\alpha$ emission extends down to $< 1.9 r_g$, and if the line comes from an accretion disk truncated at the innermost stable circular orbit this indicates a black hole spin of $a > 0.95$.

5.  The highly ionized absorber giving rise to the H- and He-like iron absorption lines in the Chandra HEG spectrum may be either intrinsic to the source or serendipitously located along the line of sight.  In the former case it must be outflowing from \mcg6 at $2.0^{+0.7}_{-0.9} \times 10^3 \kmps$, whereas in the latter is would have a velocity $0.3^{+0.7}_{-0.9} \times 10^3 \kmps$ relative to the Galaxy.  If the absorber is intrinsic to the source, it may be a highly ionized, high velocity component of the warm absorber.  This high ionization component would dominate the energy budget of the outflow and account for a significant fraction of the outflowing mass.

Future observations of the iron K band with Astro-E2 will determine the velocity shift of the H- and He-like absorption lines with greater precision and help determine the location and nature of the highly ionized absorbing gas, and place stronger constraints on ionized partial covering models.  Such observations can also be used to search for multiple velocity components of absorption lines and to look for plasma diagnostics associated with the warm absorber.

\section{Acknowledgments}

We thank John Houck and Roderick Johnstone for their help with computer software, Simon Vaughan for the XMM-Newton spectrum and Barry McKernan and Sarah Gallagher for helpful discussions and the anonymous referee for helpful comments and suggestions.  Support for this work was provided by NASA through the Smithsonian Astrophysical Observatory (SAO) contract SV3--73016 to MIT for support of the Chandra X-ray Center which is operated by the SAO for and on behalf of NASA under contract NAS8--03060.  JCL thanks and acknowledges support from the Chandra fellowship grant PF2--30023 -- this is issued by the Chandra X-ray Observatory Center, which is operated by SAO for an on behalf of NASA under contract NAS8--39073.

\bibliographystyle{apj}
\bibliography{apj-jour,bibliography}

\end{document}